# TOWARDS LLM-GENERATED EXPLANATIONS FOR COMPONENT-BASED KNOWLEDGE GRAPH QUESTION ANSWERING SYSTEMS

Dennis Schiese[1], Aleksandr Perevalov[1], Andreas Both[1,2]
[1]*HTWK Leipzig*
*Karl-Liebknecht-Straße 132, Leipzig, Germany*
[2]*DATEV eG*
*Paumgartnerstr. 6 - 14, 90429 Nuremberg, Germany*

**ABSTRACT**

Over time, software systems have reached a level of complexity that makes it difficult for their developers and users to explain particular decisions made by them. In this paper, we focus on the explainability of component-based systems for Question Answering (QA). These components often conduct processes driven by AI methods, in which behavior and decisions cannot be clearly explained or justified, s.t., even for QA experts interpreting the executed process and its results is hard. To address this challenge, we present an approach that considers the components' input and output data flows as a source for representing the behavior and provide explanations for the components, enabling users to comprehend what happened. In the QA framework used here, the data flows of the components are represented as SPARQL queries (inputs) and RDF triples (outputs). Hence, we are also providing valuable insights on verbalization regarding these data types. In our experiments, the approach generates explanations while following template-based settings (baseline) or via the use of Large Language Models (LLMs) with different configurations (automatic generation). Our evaluation shows that the explanations generated via LLMs achieve high quality and mostly outperform template-based approaches according to the users' ratings. Therefore, it enables us to automatically explain the behavior and decisions of QA components to humans while using RDF and SPARQL as a context for explanations.

## 1. INTRODUCTION

In times of rapid development of Artificial Intelligence (AI) applications, where new ones are created daily and the majority of them have become indispensable in everyday life. However, many users do not have a certain basic understanding of this technology's behavior. In a study conducted in 2021, only 28% of respondents trusted AI systems (Gillespie et al. 2021). Nevertheless, more people are in favor of further development and research into AI than are against it (Gillespie et al. 2021). The ability to explain the decisions of AI-driven systems is an important step in addressing the lack of trust among its users. In addition, developers and researchers would also benefit from better traceability (Rosenfeld & Richardson 2019).

Nowadays, many developed software systems are component-based systems, i.e., they are composed of multiple blocks isolated from each other (e.g., web services, libraries, packages). A number of these systems follow the orchestration pattern (e.g., Figure 1a). Although the way how the components are called is transparent, their internal behavior is usually encapsulated. This lack of transparency is particularly increased by the use of AI models (e.g., language models) within the components. This limits the traceability of the behavior of the system and increases the need for explainability. However, we follow here the hypothesis that due to their flexibility, component-based systems offer a decisive advantage in terms of explainability compared to monolithic systems, as different stages of a process can be considered separately, allowing for more detailed explanations. In general, the explanation of such systems typically involves providing a natural-language text, e.g., for clarifying the processes. Process explanation is central to AI methods such as

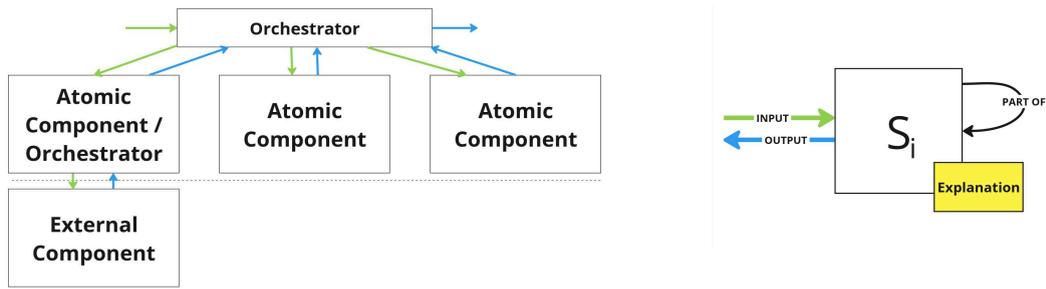

(a) Exemplary orchestrated system. The green (input) and blue (output) lines represent the data flows. The external component below the dotted line, represents an external component, whereas the parent component acts as an atomic component and orchestrator at the same time.

(b) General, generalist component model with one input and one output stream. The Part-Of label represents the possible aggregation of components of this type. The yellow explanation box indicates the possibility of explaining this component.

**Figure 1:** Example of a component-based system and generalized component model

machine learning and deep learning, with existing approaches such as (Adadi & Berrada 2018, Barredo Arrieta et al. 2020, Burkart & Huber 2021, Haar et al. 2023). Data explanation, on the other hand, aims to make system outputs understandable and transparent by explaining their individual components.

In this paper, we present an approach for creating explainable component-based Question Answering (QA) systems, followed by a case study where we utilize the semantics of the internal data flows to generate explanations with a rule-based and generative approach. Due to the lack of reference works and datasets, we won't be able to compare our results with those. For the introduction of such a dataset, we will use the Qanary framework and derived QA systems. Here, we consider the data flows of components integrated into a component-based QA system. The used QA framework is well-prepared for this approach, as it already explicitly represents the components' input data as SPARQL[1] queries (i.e., the component requests data while using SPARQL from a central process memory) and the output data as RDF[2] triples. Hence, for all components, the data flow is transparent. To validate our approach, we implement a template-based setting and one based on Large Language Models (LLM-based) for verbalizing the data flow (i.e., converting to natural language). Both settings are compared with each other using quantitative (automatic) and qualitative (human experts) settings. We derive the following *research questions*: ($RQ_1$) Can the data flow of a component be verbalized for the purpose of explaining the component's behavior?; ($RQ_2$) What is the quantitative and qualitative difference between template-based and LLM-based prompts? The paper is organized as follows. In Section 2, we give an overview of the previous work in this field. Section 3 defines the approach for explainable QA systems, and presents technical details on the implementation. The experimental setup and evaluation are presented in Section 4. We conclude the paper in Section 5.

## 2. RELATED WORK

The concept of *explainability* is inherently abstract due to the lack of a consistent definition. It was first introduced in the field of AI in 1984 in the context of software systems (Confalonieri et al. 2021). In recent years, different AI approaches have therefore emerged, aiming to explain machine- and deep-learning algorithms (Burkart & Huber 2021, Haar et al. 2023, Pope et al. 2019). While component-based QA systems may not inherently be AI-based or black boxes, the presence of AI components however justifies their consideration in terms of explainability.

---

[1] https://www.w3.org/TR/sparql11-overview/
[2] Resource Description Framework, cf. https://www.w3.org/RDF/

*Verbalizing SPARQL queries*, which can also serve to explain them, have been considered in the past, especially in the reverse direction. This consideration is interesting in the context of answering questions over linked data, but not relevant to this work. Concerning the translation of SPARQL into natural language, two popular and concrete approaches are SPARTIQULATION (Ell et al. 2015) and SPARQL2NL (Ngonga Ngomo et al. 2013). A more general approach, the explanation of structured data (e.g., SQL), is mentioned in (Koutrika et al. 2010). Regarding SPARTIQULATION and SPARQL2NL, they produce respectable results, yet the authors note room for improvement, especially with complex queries. Given this limitation and its complexity, unsuitable for the purposes of this paper, an alternative approach was chosen for subsequent considerations.

The second challenge in the scope of explanation generation is the *verbalization of grounded*[3] *RDF triples*. Although there is limited research on this topic, one approach that leverages the semantic strength of RDF is presented in (Sun & Mellish 2007). In this approach, the authors rely on an earlier study, in which they found that a large number of predicates can be classified into RDF triples (Mellish & Sun 2006). While this approach produced good results, it is only interesting as a potential fallback solution, given the risk of misinterpretation due to ontology and domain dependency (Bouayad-Agha et al. 2014).

An alternative approach to generating explanations, mentioned earlier, uses generative AI and, more specifically, *Large Language Models (LLMs)*. LLMs provide a straightforward way to generate explanations from different data sources due to their ability to generate text. The ability to fine-tune such models on instructions can further increase their potential. To evaluate the potential utility of such a fine-tuned model, we rely on the currently most powerful and promising generalized LLMs GPT-3.5 and GPT-4 from OpenAI[4] due to their generally good performance regarding the generation of text from structured data (Li et al. 2024, Ribeiro et al. 2021, Chen et al. 2020).

Nowadays, many different *Question Answering frameworks* exist, component-based or monolithic. With regard to explainable, component-based QA systems, only the approach QA2Explanation has been found. It was proposed by Shekarpour et al. (Shekarpour et al. 2020). This approach utilizes the Frankenstein approach[5] (Singh, Radhakrishna, Both, Shekarpour, Lytra, Usbeck, Vyas, Khikmatullaev, Punjani, Lange, Vidal, Lehmann & Auer 2018, Singh, Both, Sethupat & Shekarpour 2018) – an extension of the component-based Qanary framework (Both et al. 2016) – to provide valid explanations for each pipeline stage (i.e., processing steps). By augmenting the framework with classifiers, they return expected success for components and select suitable templates. Qualitative analysis of these explanations revealed that interviewees accepted them when provided, enhancing confidence in the QA process and underscoring the value of data-driven explanations for traceability. However, the authors also emphasize that the explanations depend on various factors from other disciplines, such as social science and cognitive decision-making. This reiterates the previously mentioned challenge associated with the term *explainability*. Similar results can also be found in (Hoffman et al. 2017). From the existing QA frameworks, only the Qanary framework offers the functionality required to answer our research question, as it is a component-based QA framework that manages the components as web services and explicitly represents the input and output data.

## 3. APPROACH AND IMPLEMENTATION

As a foundation for explainable QA systems, we present a concept that is aimed at systems following the orchestration pattern (cf. Figure 1a). Therefore, we define a model capable of representing the intended information as general and extensible as possible. When examining component-based systems, it becomes evident that the individual components can be considered for further analysis. By focusing on the most relevant aspects in terms of components' explainability (i.e., component input and output) in an orchestrated-based system, we derived the component model depicted in Figure 1b, where each component might integrate other components (expressed by the part-of relation). In addition, like the orchestrated system in Figure 1a, the general component model contains two data streams – *INPUT* and *OUTPUT* – and an explanation of the component's behavior for the considered specific execution (i.e., a particular process). We

---

[3] Grounded or explicit triples are triples without variables, cf. https://www.w3.org/TR/rdf-sparql-query/

[4] cf. https://platform.openai.com/docs/models

[5] cf. https://github.com/WDAqua/Frankenstein

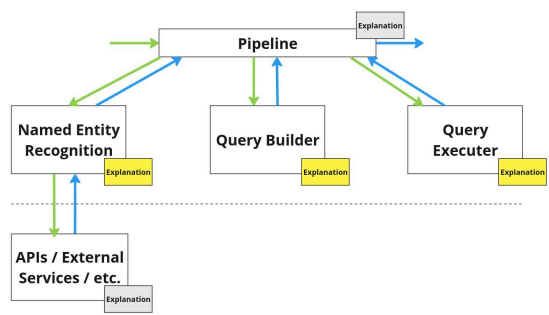

```
INPUT Template: "All Annotations of the type
AnnotationOfInstance has been requested, which
inherits the date of its creation, the URI for
the origin question, a knowledge-graph resource
as well as the correlating start and end position
for the found entity. If provided, a score is
fetched too."
OUTPUT Template: "The component ${component} has
added ${numberOfAnnotations} annotation(s) to the
graph ${graph} [...]: 1. on ${annotatedAt} &{with
a confidence of ${score}}& and the resource
${hasBody} 2. [...]"
```

(a) Exemplary Qanary system following Figure 1a. Grayed-out explanation boxes indicate that the explanation of this component is not within the scope of this work.

(b) Input/Output data templates: The input data templates don't contain placeholders, as they're pre-written and selected depending on the data requested. The output template, however, contains placeholders ${...} and &{...}, with the latter being a conditional placeholder. Each type of data has a different template.

**Figure 2:** Exemplary Qanary system and Template for both data flows

apply this concept to concrete QA systems that are built using the *Qanary framework*[6] (cf. Section 2), as these are orchestrated systems, comprising a pipeline (orchestrator) and (atomic) components. These components are web services following a particular scheme to transfer data autonomously (after being activated in a process):
- They fetch the required data from a centralized knowledge base (a triplestore) via SPARQL. Hence, the input is represented by a SPARQL SELECT query.
- After finishing the computation, each Qanary component will store the data in the process memory to make it accessible to follow-up components. Such created data is represented as grounded RDF triples and can be traced directly to the created components using Qanary-specific metadata.

Hence, the task is to verbalize the specific input data (SPARQL INSERT query) and output data (grounded RDF triples) to reflect the components' behavior regarding a user-demanded QA process identified by the question that was the starting point of the specific QA process (i.e., one that was previously executed). These explanations are to be generated in two ways: First, through a rule-/template-based approach, and second, with the use of an LLM.

**Template-based generation of component's explanations** For the rule-based approach, an external Java service[7] has been implemented to generate explanations based on the defined datasets. The *input data* is explained through fixed templates without placeholders. These templates are defined once for all central[8] available SPARQL queries. In a concrete case, the appropriate template for the executed SELECT query is chosen based on the requested data item (call annotation in the Qanary terminology). While this method offers a straightforward foundation, it is constrained in the long run as new queries require a new manual explanation. The explanation of the *output data*, specifically the RDF triples, also employs a template-based approach. In contrast to the input data, where only the graph within the SPARQL query differs, the datasets here are highly variable. Depending on the component, the question, or the existing data in the triplestore, the data changes accordingly. Consequently, placeholders are used in the templates that utilize the values from the dataset. The datasets also contain both common (e.g., metadata) and different predicates across annotation types. This circumstance necessitates a two-part explanation, which is structured as follows: (1) The prefix: Consists of the executed component and the number of data items (recognizable by different IDs) that were sent from the component to the QA system (stored in a knowledge graph); (2) The actual output data items: Explanation of each data item. The template for each item is selected based on the corresponding data type. For each item, additional pieces of information are provided (e.g., time and confidence score).

---

[6] cf. https://github.com/WDAqua/Qanary
[7] cf. https://github.com/WSE-research/qanary-explanation-service
[8] The Qanary commons package defined all used SPARQL queries in a central place:
https://github.com/WDAqua/Qanary/tree/master/qanary_commons

Following this structure, an explanation (with placeholders) looks like illustrated in Figure 2b. The explanations generated this way thus represent a rule- and template-based approach that demonstrates the applicability of the concept on the one hand and forms the basis for further considerations on the other.

**LLM-based generation of component's explanations** As the template-driven approach has several downsides, e.g., high costs for maintaining or extending templates and limitations regarding generalization (i.e., non-automated explanation capability for new annotation types), we also investigated the ability of LLMs to verbalize our data flows. To create LLM-based explanations, we used the prompt templates shown in Figure 3a (output data) and 3b (input data).

## 4. EXPERIMENTS AND EVALUATION

In this section, we begin by grouping concrete Qanary components into "component sets", which serve to generalize the various existing data types[9]. Therefore, the notation (1) $C_1,..., C_n$ represents the used components, and (2) $O_1,..., O_n$ stands for the different (output) data types the components produce. In contrast, the input data (SPARQL SELECT queries) are referenced as $I_1,...,I_6$[10]. In our experiments, the following components (from the Qanary framework) were used to produce the following data types:

1. $O_1$: $C_1$=NED-DBpediaSpotlight, $C_2$=NED Dandelion, $C_3$=NED-Ontotext, $C_4$=NED-TagMe
2. $O_2$: $C_5$=NER-DBpediaSpotlight, $C_6$=TagMeNER, $C_7$=TextRazor, $C_8$=DandelionNER
3. $O_3$: $C_9$=REL Python Falcon, $C_{10}$=DiambiguationProperty-OKBQA
4. $O_4$: $C_{11}$=SINA, $C_{12}$=QAnswerQueryBuilderAndQueryCandidateFetcher, $C_{13}$=PlatypusQueryBuilder

Subsequently, we intend to evaluate the experiments in qualitative and quantitative terms. It should be noted that the quantitative evaluation will be based on a specifically derived formula and will pertain only to the latter output-data explanation experiments, not the template-based explanations, which serve as the baseline in this context. With regard to the qualitative evaluation, we recognize the problem of a generally applicable evaluation scheme for explanations (Adadi & Berrada 2018, Anjomshoae et al. 2019). That's why we concentrated on two aspects that aim to encompass the two dimensions of explanation evaluation as described by Leilani et al. (Gilpin et al. 2018), namely interpretability and completeness. Thus, we use 'correctness' to assess completeness and 'usefulness' to assess interpretability.

The evaluation of the **baseline explanations** and all subsequent experimental explanations has been carried out by three experts, each with a research background in the field of Question Answering and Linked Data. However, in order to get feedback from a wider audience and gain further insights, we've also implemented and published a demo with feedback functionality (Schiese et al. 2024). The result shows the quality of the template-based explanations for both types: input-data explanations and output-data explanations. Furthermore, the quality was rated on a 5-point Likert scale (with "5" being the best rating). The evaluation of the *input-data explanations* was based on the separated metrics: *correctness* and *usefulness*. For both metrics, the average was 3.67, whereas the majority was rated with 4. This in turn means that the explanations seem to be 73.4% correct. In terms of their usefulness, the experts consider them to be rather useful. The evaluation for the *output-data explanations* is based on the combination of the correctness and usefulness metrics. This is due to the fact that the values are based on factual information and therefore, they're less semantically influenced by the templates. Here, the majority of ratings was 4, too. On average, these were rated at 3.704. Since we can assume that the accuracy of the factual data is probably 100%, it is reasonable to conclude that these explanations are just as useful to the experts as the explanations of the input data.

---

[9] In Qanary, these different data types are represented by different annotation types
[10] The corresponding SPARQL queries are accessible in the online appendix, cf. https://doi.org/10.6084/m9.figshare.27079687

```
Given the following context:                          Given the following context: Here, we
Here, we consider the data of a Question Answering    consider the data of a Question Answering
system. The data describes the outcome of this system.system. The data describes a SPARQL query. As
As a user I'd like to understand what happened inside a user I'd like to understand what the query
that particular component. For this purpose a         means and does. For this purpose a
(text-based) explanation has to be computed.          (text-based) explanation has to be computed.

For example, the following explanation was created for Here's an example explanation:
the question "<QUESTION_ID_EXAMPLE>" from the given
raw data.                                             The query:
The example explanation:                              ${EXAMPLE_QUERY}
<EXAMPLE_EXPLANATION>                                 The example explanation:
Given raw data:                                       "${EXAMPLE_EXPLANATION}"
<EXAMPLE_RDF_DATA>                                    Now explain the following query, used by the
Now, create an explanation for the following RDF data:component "${COMPONENT}":
<TASK_RDF_DATA_TEST>                                  ${TEST_QUERY}
Don't introduce your answer and only return the
result.                                               Don't use more than 3 sentences.
```

(a) Prompt for output data explanations      (b) Prompt for input data explanations

**Figure** 3: Prompt templates for input and output data.

All **LLM-based experiments** were performed with the generalized OpenAI models[11]. The models used were GPT-3.5 and GPT-4[12]. In addition, we used vanilla LLM configurations (i.e., no fine-tuning) due to the lack of a training dataset and the Chat Completions API without modified parameters[13]. The question set consisted of the QALD-10[14] dataset with 394 questions. Finally, the used prompts are displayed in Figures 3a and 3b.

**Evaluation approach.** As with the template-based explanation evaluation, these experiments were also evaluated qualitatively using the metrics and the questionnaire shown in Fig 4b. In the case of the output-data explanations, and as pointed out earlier, we carried out an additional quantitative analysis. Due to the variability of the RDF triples, which depended on factors such as the question, the component used, and the data available in the knowledge graph, we were faced with a significantly greater variety of data. This diversity is also reflected in the variance of the explanations as they utilize this data. Therefore, generating these explanations using LLMs required more adaptation, as more than just NLG capabilities were needed. In a preliminary study, we then found that the data referenced in the explanations were misused to varying degrees, which led us to decide on a quantitative evaluation. As a basis for this evaluation, we derived the following formula:

$$Q_E = Rating_{Prefix} + ((\sum_{i=1}^{A_{Ann}} Q_{Ann_i})/A_{Ann})$$

There, the quality of an experiment ($Q_E$) is a composition of the prefixes'[15] rating ($Rating_{Prefix}$) and the average rating of all annotations (second part of the sum), whereas $A_{Ann}$ represents the number of explainable annotations. A depreciation of a rating took place: (1) For the prefix in the case of incorrect recognition: (a) of the component, or (b) of the number of annotations; (2) For the annotation on (a) missing values (multiple occurrences only evaluated once), or (b) incorrect values (multiple occurrences only evaluated once). An exemplary evaluation following this formula is shown in Figure 4a.

---

[11] cf. https://openai.com/
[12] GPT-3.5 models: gpt-3.5-turbo / gpt-3.5-turbo-16k, GPT-4 model: gpt-4-0613
[13] The parameters are listed in the reference documentation for the Chat Completions API: cf. https://platform.openai.com/docs/api-reference/chat/create. Only the message and the model were passed.
[14] cf. https://github.com/WSE-research/QADO-datasets/releases/tag/v0.6.0
[15] The proposed output-data explanations can be divided into a prefix and the list of annotation explanations. The prefix includes information about the number of annotations and the component. This data was part of every dataset for the output-data explanations.

(a) Comparison of template-based vs. generative explanations: The yellow prefix includes the component and annotation count, while the orange part explains the annotations. The generative explanation incorrectly adds a score (annotationId) and hallucinates a resource, omitting entity position. The annotation, not the prefix, is depreciated due to these errors, leading to $Q_E=3+1=4$.

(b) Definitions of evaluation tasks and metrics used in the expert evaluation include explanations of both input and output data. Metrics for output were combined since factual information (RDF triples) was already covered in the quantitative analysis.

**Figure** 4: Evaluation: Example of output-data explanation and rating scheme.

# Results

In the following, all results are displayed and briefly discussed. The raw data for each experiment, as well as the evaluation results, are accessible in the online appendix[16].

**Input data explanations** For the input data, explanations for 6 ($I_1$ to $I_6$) different SPARQL queries were generated. Zero-, one-, and two-shot experiments were carried out two times for each of these queries (one each with the GPT-3.5 and GPT-4 models). Together with the template explanations, there were 42 explanations. Firstly, all generatively computed explanations achieved better results than the template-based ones. Furthermore, the differences between zero-, one-, and few-shot approaches seem to be small, but more examples never worsen the results. Regarding the used GPT models, it could not clearly be determined whether one was better than the other. In particular, GPT-3.5 produced better results for usefulness in zero and one-shot experiments while GPT-4 did so for correctness in one and two-shot examples as well as for usefulness in the two-shot experiments (cf. Table 1).

**Quantitative (output data explanations)** The results of the quantitative evaluation were analyzed as part of the study according to the 1- and few-shot experiment series (Table 2). In addition to anomalies, possible correlations (according to the Pearson correlation coefficient) between the overall quality and the number of annotations were also examined. For the 1-shot experiment series (see Table 2), the test datatype $O_3$ turned out to be the best with the highest average ratings. It was followed by $O_1$, $O_4$, and $O_2$. For all except $O_1$, experiments performed best when the example and test data type matched. In general, this is no surprise, because generative AI usually works best when the model is provided by concrete examples that only need to be repeated. Running the prompts with the GPT-4 model led to a significant improvement in the results regarding the average values in each series of experiments. In this case, type $O_3$ achieved the highest possible average score in two cases and thus performed best, again. Consequently, an optimized recognition and processing of grounded RDF triples can be stated. Based on these results, the 2-shot test series was carried Table 1: Qualitative expert evaluation for both data types Ii and Oi. The displayed ratings represent the average quality across all experiments for the given setting. For the output data explanations, no zero-shot examples were considered. The same applies to 2-shot experiments for GPT-4. For each metric, the best rating is highlighted (per row).

---

[16] Raw data and evaluation results: https://doi.org/10.6084/m9.figshare.27079687

|  |  | Template | Zero-shot prompt (LLM) | | | One-shot prompt (LLM) | | | Two-shot prompt (LLM) | | |
| --- | --- | --- | --- | --- | --- | --- | --- | --- | --- | --- | --- |
|  |  | - | GPT-3.5 | GPT-4 | Δ | GPT-3.5 | GPT-4 | Δ | GPT-3.5 | GPT-4 | Δ |
| $I_i$ explanations | Correctness | 3.67 | 4.50 | 4.50 | ±0.00 | 4.50 | 4.56 | +0.06 | 4.50 | **4.72** | +0.22 |
|  | Usefulness | 3.67 | 3.83 | 3.78 | −0.05 | 3.89 | 3.78 | −0.11 | 4.06 | **4.11** | +0.05 |
| $O_i$ explanations | Quality | 3.70 | ∅ | ∅ | ∅ | 3.59 | 3.71 | +0.11 | ∅ | ∅ | ∅ |

Table 2: Quantitative output-data explanation evaluation: 1-shot test series with quantitatively determined average values for each experimental setting (Exemplary output, and output to be explained). One experimental setting consisted of 50 experiments. The highlighted values represent the best result for each column and both used models.

| Evaluation | | Tested data | | | | | | | | | | | | | |
| --- | --- | --- | --- | --- | --- | --- | --- | --- | --- | --- | --- | --- | --- | --- | --- |
|  |  | $O_1$ | | | $O_2$ | | | $O_3$ | | | $O_4$ | | | dy | | |
|  |  | GPT-3.5 | GPT-4 | Δ | GPT-3.5 | GPT-4 | Δ | GPT-3.5 | GPT-4 | Δ | GPT-3.5 | GPT-4 | Δ | GPT-3.5 | GPT-4 | Δ |
| Example for prompt | $O_1$ | 5.29 | 5.60 | +0.31 | 4.98 | 5.02 | +0.04 | 5.84 | **6.00** | +0.16 | 5.41 | 5.84 | +0.43 | 5.38 | 5.62 | +0.24 |
|  | $O_2$ | 5.33 | 5.47 | +0.14 | 5.83 | **6.00** | +0.17 | 5.58 | 5.94 | +0.36 | 5.36 | 5.85 | +0.49 | 5.53 | 5.82 | +0.29 |
|  | $O_3$ | **5.62** | 5.76 | +0.14 | 5.02 | 5.79 | +0.77 | 5.99 | **6.00** | +0.01 | 4.97 | 5.66 | +0.69 | 5.40 | 5.80 | +0.40 |
|  | $O_4$ | 5.36 | **5.90** | +0.54 | 5.18 | 5.92 | +0.74 | 5.59 | 5.98 | +0.39 | 5.88 | **5.98** | +0.10 | 5.50 | **5.95** | +0.45 |

out (i.e., two examples are provided in the prompt). Each of these comprised 10 example type combinations, which represented the combination of all existing types without taking the order into account. The overall results are presented in Figure 5.

Starting with the consideration of the test data type $O_1$, the average rating is between $5.22 \leq Rating \leq 5.84$. It was noticeable, that the problem regarding missing scores and the $C_1$ component occurred here, too. The correlation coefficients were found to be in the range of $0.249 \leq |r| \leq 0.595$. For the test type $O_2$, the range of average ratings is considerably broader, with a minimum of 4.42 and a maximum of 5.90. Furthermore, four experiment series exhibit ratings below 5. It was observed that the combination of data types with $O_1$ resulted in sub-optimal performance. In contrast, the combinations with $O_2$ exhibited the most favorable performance. The primary reason for the inferior outcomes can be attributed to the presence of missing and wrong values within the annotation's explanation. With regard to the component exhibiting the highest performance, $C_6$ achieved a score of 6 in 66.7% of cases (i.e., an excellent result). Finally, the calculated correlation coefficients were found to lie between $|r| < 0.30$ for five cases and $|r| > 0.30$ for another five cases. Thirdly, the test series for $O_3$ was evaluated and, as with the corresponding 1-shot series, no anomalies were found. Nevertheless, the average ratings were below those of the 1-shot test series. However, the observed discrepancy precludes any conclusion regarding the relative performance of the two experiment series. It was also found that the distribution of ratings across both components used was almost identical. The correlation coefficients were in the range of $0.015 \leq |r| \leq 0.431$. Finally, the type $O_4$ was considered. It should be mentioned that the vast majority of experiments were carried out with the component $C_{11}$[17]. The result should therefore be considered in the context of this circumstance. However, it was noticeable that the series of experiments with at least one component of the type $O_1$ performed significantly better and ranged between 5.8 and 6.0 (i.e., very good results). Different combinations varied between $4.95 \leq Rating \leq 5.39$. The consideration of possible correlations and the component score distribution was omitted due to the aforementioned bias.

**Qualitative (output data explanations)** For the qualitative evaluation (cf. last row in Table 1) a total of 54 experiments were selected. Since this qualitative analysis of the explanations served primarily as a trend check for the results of the quantitative evaluation (see above paragraph), we limited the experiments used, resulting in a subset of 1-shot explanations of GPT-3.5 and GPT-4 as well as the template explanations. Thus, each experiment type contained 18 experiments. The metric was the same as for the template quality analysis. It was notable that both experiment types exhibit a comparable level of performance, with 3.59 for GPT-3.5 and 3.71 for GPT-4. Thus, even with only this subset of evaluated experiments, the tendency for good generative-created explanations is amplified.

---

[17] Over the duration of the experimentation, other components ceased to yield solutions.

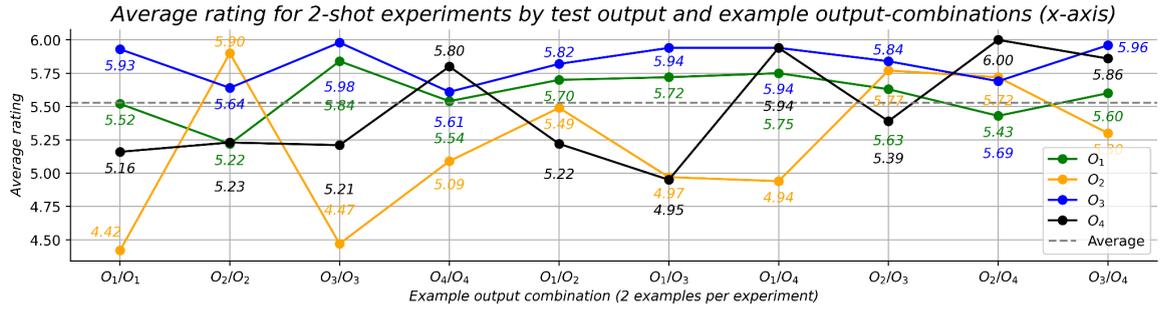

**Figure 5**: Quantitative evaluation of output-data explanations. A 2-shot test series (50 trials per setup) was conducted with two samples per prompt. The x-axis shows output-data combinations, and the dashed line marks the average of 5.53 across all experiments.

## 5. CONCLUSION

In this paper, we investigated the potential of generating explanations for Question Answering systems which are typically built as a component-based system following the orchestration pattern. These systems integrate processing steps that include AI-driven implementations, namely, natural language processing. For the sake of a generalized solution and as many AI approaches cannot be explained properly, we decided to focus in this paper on creating explanations of the observable data flows (input and output) of the components. The created explanations enable experts to receive a stepwise (component-wise) understanding of the processing of concrete component behavior and therefore the whole process. Here we used Qanary-based Question Answering systems (i.e., natural-language question processing) as a case study.

In this study, the components data flows are represented as SPARQL queries ($I_1,...,I_6$) and grounded RDF triples ($O_1,...,O_4$) that constituted the basis for these explanations. Therefore, we initially introduced a concept to represent the explanations for components and their input as well as output streams. In this second step, we demonstrated a rule-based approach to generate explanations, which served as a foundation for subsequent experimental considerations (where LLMs were used). The results of these experiments clearly showed that these explanations do not lag behind the former ones. When translating SPARQL queries, we found that LLMs produce almost correct natural-language representations of these queries. However, in our case study was no clear improvement regarding different GPT models. Regarding the explanation of the output data, here RDF triples, we focused on the adaptability between the different data types. While our quantitative evaluation showed several different results and problems, such as counting or mismatching, it also showed potential, namely the relevance of predicates. The human expert evaluation, on the other hand, clearly showed at least the same quality in comparison.

The presented approach provides a huge potential as it is not limited to the field of Question Answering. Instead, it provides a feasible method to explain the behavior of systems while establishing a semantic layer for the input and output data of each observable component in a system. As this recording can be minimally invasive, there is a wide scope of applicability. Moreover, our experiments have shown, that LLMs are well-suited for automatically generating human-readable explanations which are highly valued by the experts who typically need to understand (often growing and changing) systems while aiming for additional features, improved result quality, or must have their system audited. From this, we derive a great potential for the explanation of complex systems by using additional semantics for the data flows. Further research into the field of explainability in component-based (Question Answering) systems, utilizing a range of datasets or alternative methodologies, could prove beneficial. Additionally, following this first step, it might be worth investigating the possibility of explaining systems and their hierarchical aggregations of components to help researchers and developers understand the behavior of component-based systems.